\documentclass[acus]{JAC2000}

\def\ie{{\it\kern-2pt i.\kern-.5pt e.\kern-2pt}}  
\def\ifm#1{\relax\ifmmode#1\else$#1$\fi}
\def\ff{$\phi$--factory}  \def\DAF{DA\char8NE}  \def\f{\ifm{\phi}} 
\def\epm{\ifm{e^+e^-}} \def\gam{\ifm{\gamma}}
\def\pic{\ifm{\pi^+\pi^-}}  
\def\to{\ifm{\rightarrow}} \def\sig{\ifm{\sigma}}  
\def\up#1{\ifm{^{#1}}}  
\def\DAF{DA$\Phi$NE}  \def\dif{\hbox{d}}  
 \def\ab{\ifm{\sim}}  \def\x{\ifm{\times}}
  \def\pt#1,#2,{\ifm{#1\x10^{#2}}}
 \def\L{\ifm{{\cal L}}}  
\makeatletter
\newdimen\z@ \z@=0pt % can be used both for 0pt and 0
\newskip\z@skip \z@skip=0pt plus0pt minus0pt
\def\m@th{\mathsurround=\z@}
\def\ialign{\everycr{}\tabskip\z@skip\halign} % initialized \halign
\def\eqalign#1{\null\,\vcenter{\openup\jot\m@th
  \ialign{\strut\hfil$\displaystyle{##}$&$\displaystyle{{}##}$\hfil
      \crcr#1\crcr}}\,}
\makeatother
\usepackage{graphicx}

\begin{document}

\title{
\begin{flushright}
 \small
 Contributed paper to {\it Lepton Photon 2001}\\
 \small
 Rome, July 23-28 2001 
\end{flushright}
\vspace*{1.0truecm}
\centering MEASURING THE HADRONIC CROSS SECTION \\
AT KLOE USING THE RADIATIVE RETURN}

\author{ {\bf The KLOE Collaboration:}\\
A.~Aloisio$^g$, F.~Ambrosino$^g$, 
%{\it A.~Andryakov\mo,\hsa}
A.~Antonelli$^c$, M.~Antonelli$^c$, 
%{\it F.~Anulli\fr,\hsa} 
C.~Bacci$^l$, G.~Barbiellini$^n$, \\
F.~Bellini$^l$
G.~Bencivenni$^c$, S.~Bertolucci$^c$, C.~Bini$^j$
% ...
\\
\vspace{\baselineskip}\\
{\small $^a$Dipartimento di Fisica dell'Universit\`a e Sezione INFN, Bari, Italy}\\
{\small $^b$Institute of High Energy Physics of Academica Sinica, Beijing, China}\\
{\small $^c$Laboratori Nazionali di Frascati dell'INFN, Frascati, Italy}\\
{\small $^d$Institut f\"ur Experimentelle Kernphysik, Universit\"at Karlsruhe, Germany}\\
{\small $^e$Dipartimento di Fisica dell'Universit\`a e Sezione INFN, Lecce, Italy}\\
{\small $^f$Institute for Theoretical and Experimental Physics, Moscow, Russia}\\
{\small $^g$Dipartimento di Scienze Fisiche dell'Universit\`a e Sezione INFN, Napoli, Italy}\\
{\small $^h$Dipartimento di Fisica dell'Universit\`a e Sezione INFN, Pisa, Italy}\\
{\small $^j$Dipartimento di Fisica dell'Universit\`a ``La Sapienza'' e Sezione INFN, Roma, Italy}\\
{\small $^k$Dipartimento di Fisica dell'Universit\`a ``Tor Vergata'' e Sezione INFN, Roma, Italy}\\
{\small $^l$Dipartimento di Fisica dell'Universit\`a ``Roma Tre'' e Sezione INFN, Roma, Italy}\\
{\small $^m$Physics Department, State University of New York at Stony Brook, USA}\\
{\small $^n$Dipartimento di Fisica dell'Universit\`a e Sezione INFN, Trieste, Italy}\\
{\small $^o$Physics Department, University of Virginia, USA}
}

\maketitle

\abstract{
We present a study of the reaction \epm\to\pic\gam\ at the \f\ peak with
the KLOE detector at the \ff \hbox{ } DA$\Phi$NE.
This reaction allows us to obtain the cross section for \epm\to\pic\ 
from the \epm\ center-of-mass energy $W=m_\phi=1019$ MeV down to 
threshold, 
\ie\ $2m_\pi<M_{\pi\pi}<m_\phi$. This is called radiative return. 
The status of the analysis
and preliminary results on the invariant mass spectrum of 
the two-pion-state are presented.}
%cad]

\section{INTRODUCTION}

\subsection{Motivation}

The recent surprising results of the $g-2$ BNL experiment \cite{e821} and the 
apparent large discrepancy with expectation, about three times as large
as the electroweak correction, has renewed the interest in good measurements
of the cross section \epm\to hadrons.
The corrections to the photon
propagator due to low mass hadronic states cannot be computed, 
because QCD is not tractable in this energy regime.
It is however well known
that the correction to the muon anomaly, $a_\mu$, can be related
to $\sig(\epm\to{\rm hadrons})$ by a dispersion integral.
The error on $a_\mu({\rm hadr})$ is in fact dominated 
by the knowledge of the hadronic cross section, or equivalently
$\tau$-decay data. Quoted values for $a_\mu({\rm hadr})$ are
$(6924 \pm 62)$x$10^{-11}$ (\cite{davhoe}, using $\tau$ data) 
and $(6974 \pm 105)$x$10^{-11}$ (\cite{jeg}, using $e^+ e^-$  
data only). Taking the evaluation of Ref. \cite{davhoe}, 
theory and experiment
differ by $(426 \pm 165)$x$10^{-11}$.

\subsection{Radiative Return}

We discuss here a method to obtain $\sig(\epm\to{\rm hadrons})$, which 
employes the 
radiative process \epm\to hadrons+\gam, where the
photon has been radiated by one of the initial electrons
(positrons). This is referred to as initial state radiation or ISR
\cite{spagn}, \cite{binner}. \DAF\ typically
operates at a fixed center-of-mass energy $W=m_\phi$.
By measuring the radiative (return) process above, the hadronic cross section 
becomes accessible over the mass range $2m_\pi< M_{\pi\pi} < m_\phi$. 
The cross section 
$\dif\sigma(\epm\to{\rm hadrons})/\dif s$ can be obtained from the measurement
of $\dif\sigma(\epm\to{\rm hadrons}+\gam)/\dif s$, 
$s=W^2=M^2_{\rm hadr}$. 
The relation is given by the radiation function $H$ defined by:
\begin{equation}
s\:{\dif\sigma({\rm hadrons}+\gamma)\over\dif s} =
\sigma({\rm hadrons}) \x H(s,\theta_\gamma) 
\label{H}
\end{equation}

The function $H(s,\theta_\gamma)$ (which depends also on the polar angle
of the photon $\theta_\gamma$) needs to be known
to an accuracy better than $1\%$. 
Radiative corrections have been computed so far by different groups 
up to next-to-leading-order for the exclusive hadronic 
state $\pi^+ \pi^-$  
\cite{binner}, \cite{khoze}, \cite{german}, \cite{graz}.
We concentrated in the following on the process  
$e^+ e^- \to \rho \gamma \to \pi^+ \pi^- \gamma$, which
dominates the cross section below 
1 GeV. This channel accounts for 62$\%$ of
the hadronic contribution to $a_\mu$. 
We also limit our discussion
to events where the photon is emitted at small angle to the beams,
$0.934<\cos\theta_\gamma<0.995$ ($5^\circ<\theta_\gamma<21^\circ$) 
and is thus unobserved.
Since the event kinematics is only once constrained, large background
appears at low $M_{\pi\pi}$ due to $\pi^+\pi^-\pi^0$ and to $\mu^+\mu^-\gam$
events, both having a cross section larger than \pic\gam.
We have also begun using large angle photon events, where the 
kinematics is 4 times over constrained, three times if we ignore 
the $E_\gamma$ measurement. The analysis of this channel is
in a less advanced state but is the best way 
for studying the low dipion mass
region, which is however very important for the 
determination of $a_\mu$. 
\\
\\
The radiative return method 
has one important advantage over an 
energy scan. The systematics of the measurement
(e.g. normalization, beam energy) have to be taken into
account only {\it once} while for the energy scan they 
have to be known for {\it each} energy point. 

\subsection{Final state radiation}
Final state radiation from pions is an intrinsic background, different 
from background
processes such as muon pairs, machine background etc. and must 
be distinguished from ISR.
It is in general possible to correctly extract the relevant
\pic\gam\ contribution from the FSR background with a fit to the 
data of the cross section for both processes, after proper folding
of the detector response.
In the following we suppress FSR events by kinematic
cuts, which reduce this background to an acceptable limit. 
\\
Monte Carlo simulations\cite{ich} show, that cutting 
on $E_\gamma=(m_\phi-E_+-E_-)$ and $\theta_\gamma$, the 
energy  and polar angle
of the photon in the $e^+ e^-$ system,
effectively suppresses FSR, while 
most of the ISR events are retained. 
ISR events are peaked at large $\cos\theta_\gamma$, 
while FSR photons tend to be emitted along the pion direction
$\propto$ sin$^2\theta_\pi$.

A cut on $E_\gamma$
additionally suppresses FSR due to the fact that the decay
via the $\rho$ resonance (i.e. ISR) leads to an 
enhancement of the photon energy spectrum at $\approx 220$ MeV
at which value FSR is very suppressed. 
The cuts used in the analysis given below reduce FSR to less than $ 1\%$:
\begin{eqnarray}
 &&0.934<\cos\theta_\gamma<0.996 \label{A1} \\ 
 &&E_\gamma>10 \hbox{ MeV} \label{A2} \\
 &&55^\circ<\theta_\pi<125^\circ \label{A3} \\ 
 &&p_{\perp,\pi} > 200 \hbox{ MeV/c} \label{A4} 
\end{eqnarray}
The effective cross section with these acceptance cuts is 4.2 nb.
Note that, from the experimental point of view, it
would be much easier to perform an analysis 
with no lower boundary for $\theta_\gamma$ 
($0^\circ < \theta_\gamma < 21^\circ$). This
would increase the cross section by a factor $\simeq 4$ and
would eliminate a detector resolution effect due to the lower
angular cut. However, this is not correct in the current version
of the generator, which diverges for $\theta_\gamma \rightarrow 0^\circ$. 
A new calculation with the NLO QED corrections
taken into account  has recently been published\cite{german}
and a new version of the generator will soon be available.
This will allow to remove the lower angular cut improving
the statistical error and removing a possible source
of systematics.
\begin{figure}[t]
\centering
\includegraphics*[width=55mm]{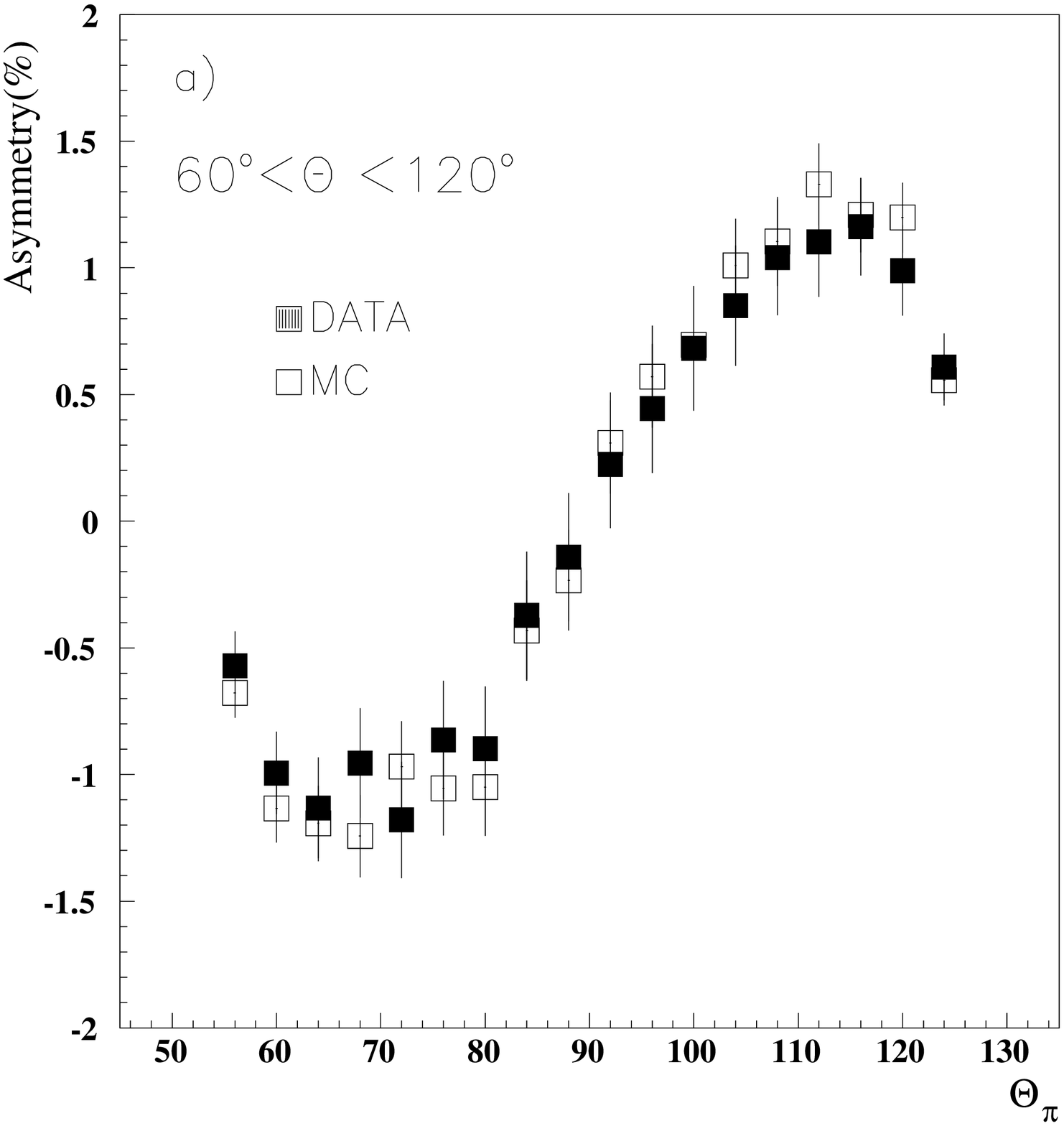}
%\vspace*{1.0truecm}
\includegraphics*[width=55mm]{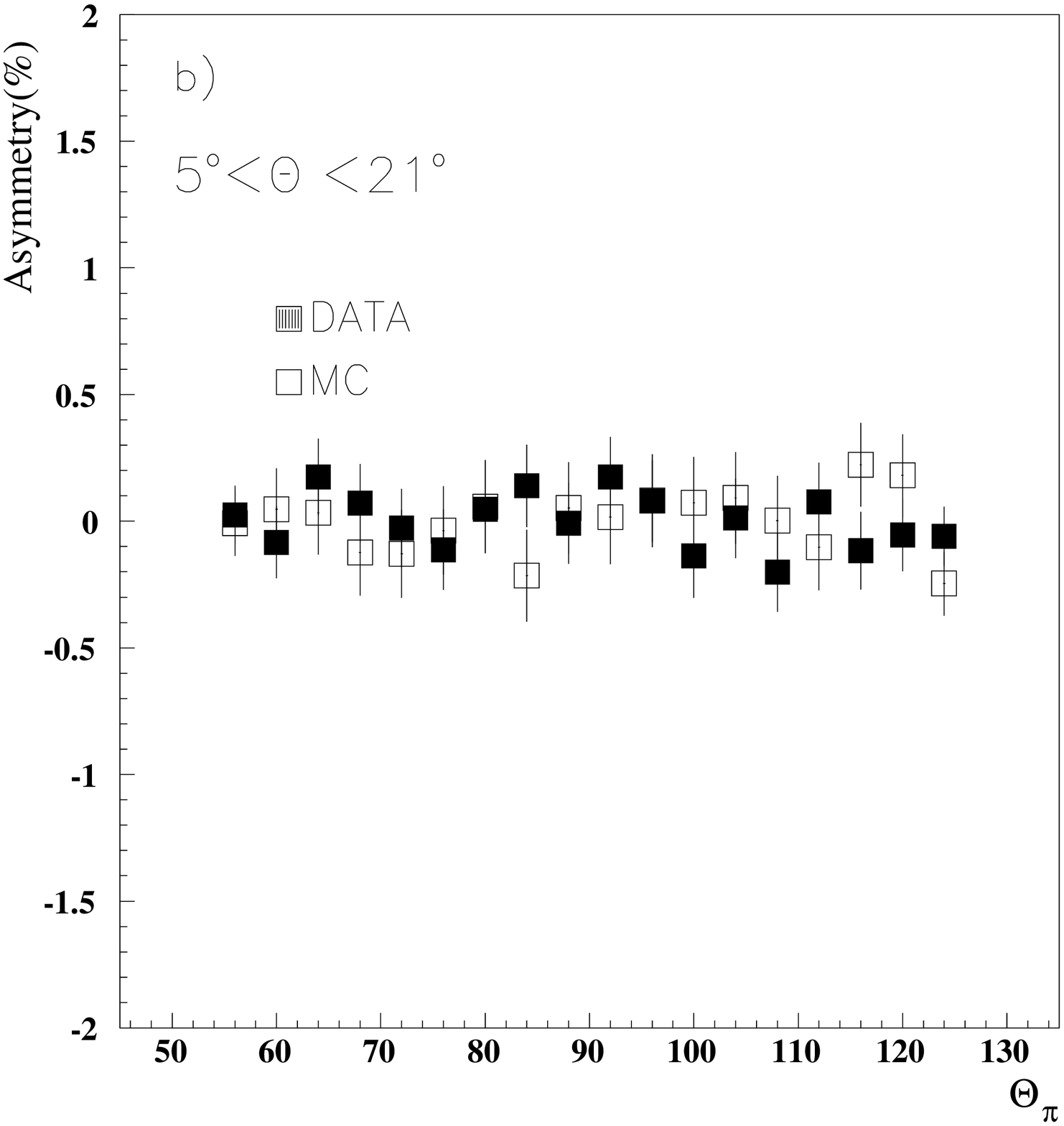}
\caption{The charge asymmetry (formula (\ref{qassym}))
is shown for 2 angular regions of the photon polar angle:
(a) $60^\circ < \theta_\gamma < 120^\circ$, where we see a sizable
effect of the charge asymmetry due to larger FSR and 
(b) $5^\circ < \theta_\gamma < 21^\circ$, where FSR and hence the
charge asymmetry are small.} 
\label{asymmetry}
\end{figure}
\\
\\
The description of FSR is model dependent and 
in the actual version of Ref. \cite{binner} a point 
like pion is assumed. The model dependence can be tested for 
$\pi^+\pi^-\gamma$ events by looking at the charge asymmetry 
of the produced pions:
\begin{equation}
A(\theta_\pi) = 
\frac{N^{\pi^+}(\theta_\pi)-N^{\pi^-}(\theta_\pi)}
     {N^{\pi^+}(\theta_\pi)+N^{\pi^-}(\theta_\pi)}
\label{qassym}
\end{equation}

This charge asymmetry 
arises from the interference between ISR and FSR and is 
therefore linear in the FSR amplitude. 
We measured the charge asymmetry for 
(a) large photon angles ($60^\circ < \theta_\gamma < 120^\circ$), 
where FSR is large, and (b) for small 
photon angles ($5^\circ < \theta_\gamma < 21^\circ$).
The results are illustrated in figure (\ref{asymmetry}) and show 
very good agreement between data and Monte Carlo,
indicating that the point like model 
describes well the process of FSR within the error bars.

\begin{figure*}[t]
\centering
\includegraphics*[width=70mm]{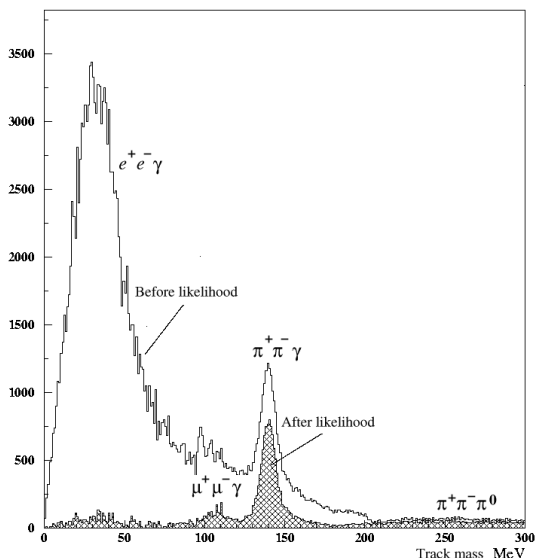}
\caption{A likelihood method has been developed to separate 
electrons from pions. The kinematic variable
track mass before and after the application of this method is shown. 
$\pi^+ \pi^- \gamma$ events are peaked at $M_{Track} = m_\pi$,
radiative Bhabha events at much smaller values. 
Other background channels are also visible 
($\mu^+ \mu^- \gamma$ and $\pi^+\pi^-\pi^0$).} 
\label{likeli}
\end{figure*}

\subsection{Studying the Pion Form Factor at threshold}

For the selection cuts described above (small angle region), 
the ISR cross section around the $\rho$ peak is highest, but 
the $M_{\pi\pi}^2$ region at threshold is suppressed by the kinematical
cuts. To avoid this, a second angular region
has been defined in which we can explore 
the invariant mass spectrum down to $\sqrt{M_{\pi\pi}^2} \simeq 300$ MeV:

\begin{eqnarray}
 &&0<\cos\theta_\gamma<0.5 \label{A6} \\ 
 &&E_\gamma > 10 \hbox{ MeV} \label{A7} \\
 &&55^\circ < \theta_\pi < 125^\circ \label{A8} \\ 
 &&p_{\perp,\pi} > 200 \hbox{ MeV/c} \label{A9} 
\end{eqnarray}

As discussed above, FSR is in average much higher at large values 
of $\theta_\gamma$, but does not play a role anymore for
small values of $M_{\pi\pi}^2$, which we actually want to study with these 
new cuts.

\section{EVENT SELECTION}

In this chapter we present the event selection
for the measurement of the $\pi^+ \pi^- \gamma$ final state. 
After a very brief description of the KLOE detector, 
the selection algorithm for this signal is presented. 

\subsection{The KLOE detector}

KLOE \cite{kloe} is a typical $e^+ e^-$ multiple purpose detector 
with cylindrical geometry,
consisting of a large helium based drift chamber (DC,
\cite{dc}),
surrounded by an electromagnetic calorimeter
(EmC, \cite{emc}) and a superconducting magnet ($B=0.6$ T). 
The detector has been
designed for the measurement of $CP$ violation in the neutral kaon system,
i.e. for precise detection of the decay products of $K_S$ and $K_L$.
These are low momenta charged tracks 
($\pi^\pm, \mu^\pm, e^\pm$ with a momentum range from $150 $ MeV/c 
to $270 $ MeV/c) and low energy photons (down to $20$ MeV). 
\\
The DC dimensions ($3.3$ m length, $2$ m radius), 
the drift cell shapes ($2$x$2$ cm$^2$ cells for the inner $12$ 
layers, $3$x$3$ cm$^2$ cells for the outer $46$ layers) and the choice of 
the gas mixture ($90\%$ Helium, $10\%$ Isobutane; $X_0 = 900$ m) 
had to be optimized for the requirements prevailing at  
a $\phi$ factory. The KLOE design results in a very good momentum 
resolution: $\sigma_{p_{\bot}} / p_{\bot} \leq 0.3\%$ at high
tracking efficiencies ($> 99\%$).
\\
The EmC is made of a matrix of scintillating fibres embedded 
in lead, which guarantees a good energy re\-solution 
$\sigma_E / E = 5.7 \% / \sqrt{E({\rm GeV})}$ and excellent 
timing resolution $\sigma_t = 57 {\rm ps} / \sqrt{E({\rm GeV})} \oplus 50$ ps.
The EmC consists of a barrel and two endcaps  
which are surrounding the cylindrical DC; this gives a 
hermetic coverage of the solid angle ($98\%$). 
However, the acceptance of the EmC below $\approx 20^\circ$ is 
reduced due to the presence of quadrupole magnets 
close to the interaction point and does not allow to 
measure e.g. the photon of $\pi^+ \pi^- \gamma$ events with low 
$\theta_\gamma$ angles (as required for FSR suppression). 
\\
\\
It will be shown in the following, that an efficient
selection of the $\pi^+ \pi^- \gamma$
signal is possible, without requiring 
an explicit photon detection. 
The relatively simple signature of the signal
(2 high momentum tracks from the interaction point) and  
the good momentum resolution of the KLOE tracking detector
allow us to perform such a selection. 
%%%The signature of this signal 
%%%(2 high momenta tracks from the interaction point) 
%%%is relatively easy compared to a typical 
%%%$K_L$ decay (2 low momenta tracks with a decay 
%%%vertex in the DC volume far away from the interaction point). 

\subsection{Selection Algorithm}

The $\pi^+ \pi^- \gamma$ events are selected using the following
4 steps. The selection is based on the measurement of the charged 
pion tracks by the DC, while the photon is not 
required to be detected in the EmC. 
Calorimeter information
is however used for the $\pi/e$-separation (likelihood method).  
\\
\\
i) Charged vertex in DC. 
We require 1 vertex in the DC with 
2 associated charged tracks 
close to the interaction point: $\sqrt{(x_V^2 + y_V^2)} \le 
8$ cm and $| z_V | \le 15$ cm.

ii) Likelihood Method for $\pi / e$ - separation.
A fraction of radiative Bhabha events $e^+ e^- \gamma$ enters the 
kinematical selection (see next selection cut), 
giving rise to a non negligible background. In order to reject those 
events, a likelihood method has been worked out for an effective
$\pi/e$-separation. The method is based on 
the shape and energy deposition of the 
EmC clusters produced by the charged tracks 
and has been developed using independent control samples for the 
pion information ($\pi^+\pi^-\pi^0$ events) and for the
electron information ($e^+ e^- \gamma$ events).  
$98\%$ of all $\pi^+ \pi^- \gamma$ events are selected if at
least one of the two tracks has been identified as a pion by
the likelihood method. In figure (\ref{likeli}) 
the effect of the likelihood method is demonstrated in the track mass 
($M_{Track}$) distribution\footnote{The $M_{Track}$ variable 
is obtained applying the 4-momentum-conservation and resolving 
it for the particle mass. }. $\pi^+ \pi^- \gamma$ events
are peaked at $M_{Track} = m_\pi$, radiative
Bhabha events at smaller values. 

iii) Kinematic Cut: $130.2 \hbox{ MeV} < M_{Track} < 149.0 \hbox{ MeV}$.
We perform a $\pm 9.6$ MeV ($\pm 2\sigma$)
cut on the kinematical variable $M_{Track}$, which is peaked 
at $M_{Track} = m_\pi$ for $\pi^+ \pi^- \gamma$ events 
(see figure (\ref{likeli})).  Background of  
$\pi^+\pi^-\pi^0$ events (with $M_{Track}$ mostly $>150$ MeV), 
$\mu^+ \mu^- \gamma$ (with $M_{Track}$ peaked at $m_\mu$) and
the bulk part of radiative Bhabha events $e^+ e^- \gamma$ 
($M_{Track}< 100$ MeV) are mostly rejected 
(see subchapter 3.1 on background). This cut assumes a
$\pi^+ \pi^- \gamma$ final state. 

iv) Acceptance Cuts.
The missing momentum of the 2 accepted charged tracks is 
calculated and associated with the photon under the 
assumption of a $\pi^+ \pi^- \gamma$ event:
$\vec{p}_{\gamma} =  \vec{p_{\phi}} - \vec{p_+} - \vec{p_-}$
\footnote{$\vec{p_{\phi}}$ is the $\phi$ momentum in the laboratory
system due to the beam crossing angle.}.
The acceptance
cuts of formulae (\ref{A1})-(\ref{A4}) are then applied.

\section{EVENT ANALYSIS}

The $\pi^+ \pi^- \gamma$ cross section measurement
contains the following terms:
\begin{equation}
\frac{d\sigma_{{\rm hadrons}+\gamma}}{dM_{\pi\pi}^2}=
\frac{dN_{{\rm Obs}}-dN_{{\rm Bkg}}}{dM_{\pi\pi}^2} \: 
\frac{1}{\epsilon_{{\rm Sel}}\epsilon_{{\rm Acc}}} \:
\frac{1}{\int\mathcal{L}dt}
\end{equation}

It requires the study of the various background channels 
($N_{{\rm Bkg}}$, see following subchapter), the selection 
efficiencies ($\epsilon_{{\rm Sel}}$) and the systematic effects 
due to the acceptance cuts applied ($\epsilon_{{\rm Acc}}$).
Finally the counting rate measurement has to be normalized to the
integrated luminosity $\int\mathcal{L}dt$
in order to achieve a cross section 
measurement. All these terms will be obtained from data for 
the final analysis.
\\
Preliminary results concerning $\epsilon_{{\rm Sel}}$ are
presented in the following subchapter. 
The detector resolution and the systematic effects, arising from there
have been studied in detail with MC \cite{ich}.
No limititations have been found for a high precision measurement 
at the percent level.

\subsection{Background}

{\bf $e^+ e^- \gamma$, $\mu^+ \mu^- \gamma$}\\
Radiative Bhabha events are mostly suppressed by the 
$M_{Track}$ cut (chapter 2.2 (ii)).
The remaining background due to 
events with a high value for $M_{Track}$ and due to 
electrons, which have not been 
rejected by the likelihood method, is peaked at large
$M_{\pi\pi}^2$ (above $0.7$ GeV$^2$) and corresponds to a 
contamination below the percent level in this $M_{\pi\pi}^2$ region. 
\\
$\mu^+ \mu^- \gamma$ events
are not efficiently rejected by the likelihood method, because they
release a pattern in the EmC with a signature similar to pions. 
After the cuts of formulae (\ref{A1}) - (\ref{A4}) and after 
the $M_{Track}$ cut, the remaining $\mu^+ \mu^-$ cross section 
is low ($\approx 10^{-2}$ nb), such that 
we expect only a small contamination ($<1\%$ in the high
$M_{\pi\pi}^2$ region $> 0.7$ GeV$^2$).
\\
\\
{\bf $\pi^+ \pi^- \pi^0$ }\\
An important background for our signal is the decay  
$\phi \to \pi^+ \pi^- \pi^0$ (B.R. $15.5 \%$, 
$\sigma_{\pi^+ \pi^- \pi^0}^{tot} \approx 500$ nb). 
$\pi^+ \pi^- \pi^0$ and $\pi^+ \pi^- \gamma$ events
are separated in the KLOE standard reconstruction scheme
by a cut in the 2-dimensional plane $M_{Track} - M_{\pi\pi}^2$. 
At small $M_{\pi\pi}^2$, the $M_{Track}$ values for the 2 channels 
are very similar and a part of the $\pi^+ \pi^- \pi^0$ events 
appear as background.
The $M_{Track}$ cut and the acceptance cut of formula 
(\ref{A4}) \footnote{The pion tracks have on the average a lower
momentum as $\pi^+ \pi^- \gamma$ events.} 
reject a big part of these events.
\\
In order to estimate from data the remaining contamination,
we modified the standard cut in the $M_{Track} - M_{\pi\pi}^2$ plane
by expanding the $\pi^+ \pi^- \pi^0$ selection 
region. We see then the tail of 
$\pi^+ \pi^- \pi^0$ events entering the $M_{Track}$ 
selection interval. We perform this study in bins of
$M_{\pi\pi}^2$. The $\pi^+ \pi^- \pi^0$ background is 
negligible in most of the $M_{\pi\pi}^2$ region and gives 
only a contamination at the lower end of the spectrum between
$0.3$ GeV$^2$ and $0.4$ GeV$^2$. The effective 
$\pi^+ \pi^- \pi^0$ cross section after all the selection steps
is $< 0.01$ nb. It increases considerably if we select 
$\pi^+ \pi^- \gamma$ events at larger polar angles of the 
photon\footnote{This behaviour can
be easily explained by the missing momentum
of $\pi^+ \pi^- \pi^0$ events (in this case the $\pi^0$),
which is peaked at large angles.}

\subsection{Selection Efficiencies}

We shortly summarize the various efficiencies 
which contribute to the total selection efficiency:

%\begin{figure*}[t]
%\centering
%\includegraphics*[width=90mm]{fpi_2_gamma_range_bw.eps}
%\caption{The pion form factor as a function of the $\sqrt{s}$
%of the hadronic system for small and large angles. The
%line is a Gounaris-Sakurai parametrization with fixed
%parameters. No fit has been attempted.
%} \label{dsigmadq2}
%\end{figure*}

\begin{itemize}

\item{ Trigger: The trigger efficiency has 2 contributions:
the probability of a $\pi^+ \pi^- \gamma$ event to
be recognized by the KLOE trigger 
(between $95\%$ to $99\%$ depending on $M_{\pi\pi}^2$)
and the inefficiency which arises
from a trigger hardware veto for the filtering of cosmic 
ray events. The second contribution causes an inefficiency 
for $\pi^+ \pi^- \gamma$ events at large
$M_{\pi\pi}^2$ ($30\%$ at $1$ GeV$^2$) 
which decreases with lower $M_{\pi\pi}^2$
(fully efficient at $\approx 0.7$ GeV$^2$).
These values have been obtained from data by looking at the 
individual probabilities for $\pi^+$ and $\pi^-$ 
to fire 1 trigger sector and 1 cosmic veto sector.}

\item{ Reconstruction Filter: A software filter is implemented 
in the KLOE reconstruction program for the filtering of 
non-collider physics 
events, like e.g. machine background and cosmic ray events. 
The inefficiency for $\pi^+ \pi^- \gamma$ events 
caused by this filter is $\approx 2\%$ (taken from MC). }

\item{ Event Selection (see subchapter 2.2): 
The DC vertex efficiency ($\approx 95\%$) is obtained from
Bhabha events, selected without requiring DC information.
The efficiency due to the likelihood selection 
is $\approx 98\%$ and is evaluated 
from data during the construction of the likelihood method.
The efficiency due to the $M_{Track}$ cut 
($\approx 90\%$) is evaluated from Monte Carlo at present.  }

\end{itemize}

\subsection{Luminosity Measurement}

The DA$\Phi$NE luminosity is measured by KLOE using
large angle Bhabha (LAB) events.
The effective Bhabha cross section 
at large angles 
($55^\circ < \theta_{+,-} < 125^\circ$) is still as high as $425$ nb.  
The number of LAB candidates $N_{LAB}$ are counted and 
normalized to the 
effective Bhabha cross section, obtained from Monte Carlo: 

\begin{equation}
\int\mathcal{L}dt = 
\frac{N_{LAB}(\theta_{+,-})}{\sigma_{LAB}^{MC}(\theta_{+,-})}
\cdot (1-\delta_{Bkg})
\end{equation}

Hence, the precision of this measurement depends on:
(i) the theoretical knowledge of the Bhabha scattering process 
including radiative corrections;
(ii) the simulation of the process by the detector simulation
program.
\\
\\
For the theory part we are using 2 independent 
Bhabha event generators: 
the Berends-Kleiss \cite{berends} generator, modified for 
DA$\Phi$NE in \cite{drago} and the BABAYAGA 
generator\cite{babayaga}. 
\\
We use a selection algorithm for LAB events with a reduced number of 
cuts, for which we expect a very good description by the 
KLOE detector simulation program. The acceptance region for the 
electron and positron polar angle ($55^\circ < \theta_{+,-} < 125^\circ$) is
measured by the EmC clusters produced by these tracks, 
while the energy measurement 
($E_{+,-} > 400$ MeV) is performed by the high resolution drift
chamber. Taking the actual detector resolutions, we expect
the systematic errors arising from these cuts to be well 
below $1\%$. Background 
from $\mu^+ \mu^- (\gamma)$, $\pi^+ \pi^- (\gamma)$ and 
$\pi^+ \pi^- \pi^0$ is below $1\%$ and
can be easily subtracted. All the selection efficiencies 
concerning the LAB measurement 
(Trigger, EmC cluster, DC tracking) are above $98\%$ and are
well reproduced by the detector simulation. 
\\
As a goal we expect to measure the DA$\Phi$NE luminosity 
at the level of $1\%$. The very good agreement of the 
experimental distributions ($\theta_{+,-}$, $E_{+,-}$) with
the existing event generators and a cross check with an independent
luminosity counter based on 
$e^+ e^- \to \gamma \gamma (\gamma)$, indicate
a good accuracy. However, more systematic checks (e.g. 
the effect of a varying beam energy and of a non centered
beam interaction point) are still to be done. 

\begin{figure}[h]
\centering
\includegraphics*[width=75mm]{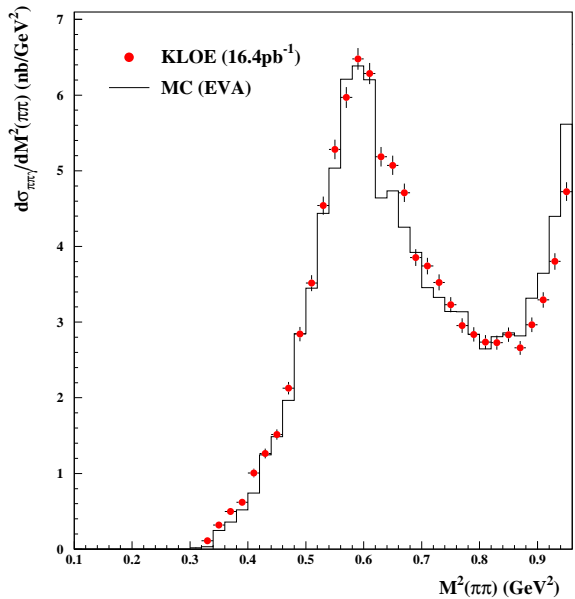}
\caption{The differential $\pi^+ \pi^- \gamma$ 
cross section as a function of the pion invariant mass.
The solid line is the prediction of the theoretical 
event generator EVA of Ref. \cite{binner}.} \label{mpipi}
\end{figure}

\subsection{Comparison with Monte Carlo}
\begin{figure*}[t]
\centering
\includegraphics*[width=90mm]{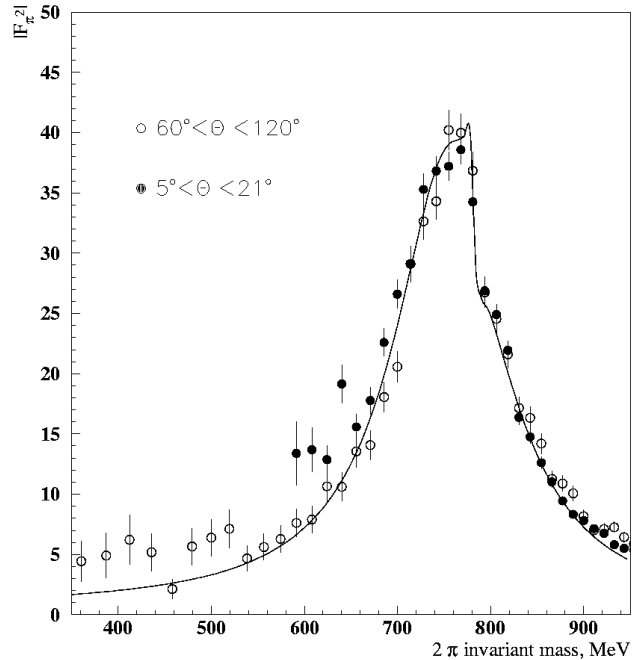}
\caption{The pion form factor as a function of the $\sqrt{s}$
of the hadronic system for small and large angles. The
line is a Gounaris-Sakurai parametrization with fixed
parameters. No fit has been attempted.
} \label{dsigmadq2}
\end{figure*}

We analyzed a data sample of 16.4 pb\up{-1}
and present in figure \ref{mpipi} the 
preliminary result for 
the dipion invariant mass spectrum $M^2_{\pi\pi}$ (small angle region)
and in figure \ref{dsigmadq2} the
pion form factor $|F_\pi|^2$.
Therefore we applied formula (\ref{H}) and write for 
$\sigma_{\rm hadrons}=\pi\alpha^2 \beta^3_\pi/(3M_{\pi\pi}^2)
\x|F_\pi(M_{\pi\pi}^2)|^2$.
In plot \ref{dsigmadq2} both the small angle
(formulae (\ref{A1})-(\ref{A4}))
and the large angle events (formulae (\ref{A6})-(\ref{A9}))
are plotted. 
The solid line is a Gounaris-Sakurai parametrization
with a fixed set of parameters and allows a 
qualitative comparison with data. 
No attempt has been made
to subtract the residual contribution of 
$\phi \rightarrow \pi^+ \pi^- \pi^0$ events 
which are present at low $M^2_{\pi\pi}$.

The statistical error of the data points in the $\rho$ peak
region is $\approx 2\%$.
The actual version of the event generator has a 
systematic uncertainty of the same size.
By comparing the data and MC distributions we 
conclude that the accuracy of our cross section
measurement is at the level of a few percent, which will be
considerably improved with the ongoing analysis. 

\section{SUMMARY AND OUTLOOK}
We presented in this paper a preliminary measurement 
$\dif\sigma(e^+ e^- \to \pi^+ \pi^- \gamma)/\dif M^2_{\pi\pi}$ 
for $M^2_{\pi\pi}<1$ GeV\up2, 
using the radiative return method. The data sample\footnote{
This corresponds to about one half of the full
data set which KLOE has taken in 2000.}, corresponding to 
an integrated luminosity of 16.4 pb\up{-1} 
shows a good agreement with the ``standard'' parametrization for
the $\rho$ peak. It also shows the possibility for KLOE
of exploring the hadronic cross section at threshold.
We conclude that the experimental 
understanding of efficiencies, background and luminosity 
are well under control. 
We further conclude that the radiative return is a competitive
method to measure hadronic cross sections.
We note again
the advantage of the radiative return. Systematic
errors, like luminosity and beam energy, enter only globally 
and do not have to be known for individual energy
points. 

For the future we plan to refine the analysis and to 
include very small angle photons. A comparison with MC will be possible 
in this case with the new
next-to-leading-order event generator \cite{german}
and will improve the precision of
the measurement due to a better systematic control of the 
fiducial volume. 
Moreover we are studying the possibility to enlarge 
the acceptance region for $\theta_{+,-}$, which increases
the kinematical acceptance of events at low $M^2_{\pi\pi}$ $,<$0.3 GeV\up2.
In order to improve on the accuracy of
$a_\mu({\rm hadr})$ and to be competitive 
with results coming from the CMD-2 experiment in 
Novosibirsk \cite{cmd2}, a final precision for this 
measurement below $1 \%$ is
needed. A statistical error on this level can be obtained with an integrated
luminosity of  \ab200 pb\up{-1} which could be collected before year end.
\DAF\ has recently achieved a luminosity 
\L$>$1.5 pb\up{-1} d\up{-1}.

We are also studying how to measure 
$\dif\sigma(e^+ e^-\to\pi^+ \pi^- + n\gamma))/\dif M^2_{\pi\pi}$,
which simply means chosing events which satisfy
appropriate kinematical constraints to reject unwanted backgrounds
and then apply precise tranverse momentum and energy unbalance for proper
inclusion of soft photons, to facilitate computation of the physical
cross section.
The radiative corrections for such a 
measurement are obtainable with high precision  
($< 0.5\%$)\cite{hoefer}. 

At low energy, $M_{\pi\pi}<600$ MeV, the $\mu^+\mu^-\gamma$ cross section
becomes larger
than the dipion cross section. It therefore becomes possibile
to measure the ratio $\sigma_{\pic\gam}/\sigma_{\mu^+\mu^-\gam}$
in which initial state radiation and vacuum polarization
effects cancel. Final state radiation corrections are still 
necessary. On the experimental side one needs only to estimate
the difference in the efficiency for the two channels rather
then the entire efficiency.

\section{ACKNOWLEDGEMENT}

We would like to thank F. Jegerlehner, J.H. K\"uhn, 
G. Rodrigo and W. Marciano for the very useful discussions and 
critical remarks during the preparation and the analysis
of this measurement.


\begin{thebibliography}{9}

%\bibitem{eidjeg}
%S. Eidelman and F. Jegerlehner, Z.Phys. C {\bf 67} (1995) 585

\bibitem{e821}
H.N. Brown {\it et al.} [E821 collaboration], 
Phys.Rev.Lett. {\bf 86} (2001) 2227

\bibitem{davhoe}
M. Davier and A. H\"ocker, Phys.Lett. B {\bf 435} (1998) 427

\bibitem{jeg}
F. Jegerlehner, [hep-ph/0104304]

%\bibitem{marc}
%W.J. Marciano and B.L. Brown, [hep-ph/0105056]

%%%\bibitem{kiril}
%%%K. Melnikov, [hep-ph/0105267]

\bibitem{spagn} 
S. Spagnolo, Eur. Phys. J. C {\bf 6} (1999) 637

\bibitem{binner}
S. Binner, J.H. K\"uhn, K. Melnikov, Phys.Lett. B {\bf 459} (1999) 279

\bibitem{khoze}
V.A. Khoze, M.I. Konchatnij, N.P. Merenkov, G. Pancheri, 
L. Trentadue, O.N. Shekhovyova, Eur. Phys. J. C {\bf 18} (2001) 481

\bibitem{german}
G. Rodrigo, A. Gehrmann-De Ridder, M. Guilleaume, J.H. K\"uhn, 
[hep-ph/0106132]

\bibitem{graz}
K. Melnikov, F. Nguyen, B. Valeriani, G. Venanzoni, 
Phys.Lett. B {\bf 477} (2000) 114

\bibitem{hoefer}
A. H\"ofer, J. Gluza and F. Jegerlehner, 
{\it Pion Pair Production with Higher Order Radiative Corrections
in Low Energy $e^+e^-$ Collisions}, 
{\bf DESY-00-163} (2001)

\bibitem{ich}
A.Denig, Proceedings to {\it DA$\Phi$NE'99}, 
{\bf Frascati Physics Series XVI} (2000) 569

\bibitem{kloe}
A. Aloisio {\it et al.} [KLOE collaboration],
{\it The KLOE Detetor, Technical Proposal},
{\bf LNF-93/002} (1993)

\bibitem{dc}
M. Adinolfi {\it et al.} [KLOE collaboration],
{\it The Tracking Detector of the KLOE
Experiment}, {\bf LNF-01/016} (2001); to be published in Nuclear 
Instruments and Methods

\bibitem{emc}
M. Adinolfi {\it et al.} [KLOE collaboration],
{\it The KLOE electromagnetic calorimeter},
{\bf LNF-01/017} (2001); to be published in Nuclear 
Instruments and Methods

\bibitem{berends}
F.A. Berends, R. Kleiss, Nucl.Phys. B {\bf 228} (1988) 537

\bibitem{drago}
E. Drago, G. Venanzoni, {\it A Bhabha Generator for DA$\Phi$NE
including radiative corrections and $\phi$ resonance}, 
{\bf INFN/AE-97/48} (1997)

\bibitem{babayaga}
C.M.C. Calame, C. Lunardini, G. Montagna, O. Nicrosini, 
F. Piccinini, Nucl.Phys. B {\bf 584} (2000) 459

\bibitem{cmd2}
R.R. Akhmetshin {\it et al.} [CMD-2 collaboration], [hep-ex/9904027]

\end{thebibliography}
\end{document}